
\font\titlefont = cmr10 scaled\magstep 4
 2
\font\sectionfont = cmr10
\font\littlefont = cmr5 
\font\eightrm = cmr8

\def\ss{\scriptstyle}
\def\sss{\scriptscriptstyle}

\newcount\tcflag
\tcflag = 0  

\ifnum\tcflag = 0 \magnification = 1200 \fi  

\global\baselineskip = 1.2\baselineskip 
\global\parskip = 4pt plus 0.3pt 
\global\abovedisplayskip = 18pt plus3pt minus9pt
\global\belowdisplayskip = 18pt plus3pt minus9pt
\global\abovedisplayshortskip = 6pt plus3pt
\global\belowdisplayshortskip = 6pt plus3pt


\def\endignore{}
\def\ignore #1\endignore{} 

\newcount\dflag
\dflag = 0


\def\monthname{\ifcase\month 
\or January \or February \or March \or April \or May \or June%
\or July \or August \or September \or October \or November %
\or December 
\fi}

\newcount\dummy
\newcount\minute  
\newcount\hour
\newcount\localtime
\newcount\localday
\localtime = \time
\localday = \day

\def\advanceclock#1#2{ 
\dummy = #1
\multiply\dummy by 60
\advance\dummy by #2
\advance\localtime by \dummy
\ifnum\localtime > 1440 
\advance\localtime by -1440
\advance\localday by 1
\fi}

\def\settime{{\dummy = \localtime %
\divide\dummy by 60%
\hour = \dummy 
\minute = \localtime%
\multiply\dummy by 60%
\advance\minute by -\dummy 
\ifnum\minute < 10
\xdef\spacer{0} 
\else \xdef\spacer{}
\fi %
\ifnum\hour < 12
\xdef\ampm{a.m.} 
\else
\xdef\ampm{p.m.} 
\advance\hour by -12 %
\fi %
\ifnum\hour = 0 \hour = 12 \fi 
\xdef\timestring{\number\hour : \spacer \number\minute%
\thinspace \ampm}}}



\def\endtitle{}
\def\title#1\endtitle{\vskip.5in\titlefont
\global\baselineskip = 2\baselineskip 
#1\vskip.4in
\baselineskip = 0.5\baselineskip\rm}

\def\endauthors{}
\def\authors#1\endauthors{#1}

\def\endabstract{}
\def\abstract#1\endabstract{\vskip .3in%
\centerline{\sectionfont\bf Abstract}%
\vskip .1in
\noindent#1}

\def\nopageonenumber{\footline={\ifnum\pageno<2\hfil\else
\hss\tenrm\folio\hss\fi}}  

\newcount\nsection 
\newcount\nsubsection 

\def\section#1{\global\advance\nsection by 1
\nsubsection=0
\bigskip\noindent\centerline{\sectionfont \bf \number\nsection.\ #1}
\bigskip\rm\nobreak}

\def\subsection#1{\global\advance\nsubsection by 1
\bigskip\noindent\sectionfont \sl \number\nsection.\number\nsubsection)\
#1\bigskip\rm\nobreak}

\def\topic #1{{\medskip\noindent $\bullet$ \it #1:}}
\def\endtopic{\medskip}

\def\appendix#1#2{\bigskip\noindent%
\centerline{\sectionfont \bf Appendix #1.\ #2} 
\bigskip\rm\nobreak} 


\newcount\nref 
\global\nref = 1 

\def\therefs{}


\def\ref#1#2{\xdef #1{[\number\nref]} 
\ifnum\nref = 1\global\xdef\therefs{\item{[\number\nref]} #2\ } 
\else
\global\xdef\oldrefs{\therefs}
\global\xdef\therefs{\oldrefs\vskip.1in\item{[\number\nref]} #2\ }%
\fi%
\global\advance\nref by 1
}

\def\listrefs{\vfill\eject\section{References}\therefs}


\newcount\nfoot 
\global\nfoot = 1 

\def\foot#1#2{\xdef #1{(\number\nfoot)} 
\footnote{${}^{\number\nfoot}$}{\eightrm #2}
\global\advance\nfoot by 1
}


\newcount\nfig 
\global\nfig = 1
\def\thefigs{} 

\def\figure#1#2{\xdef #1{(\number\nfig)}
\ifnum\nfig = 1\global\xdef\thefigs{\item{(\number\nfig)} #2\ }
\else
\global\xdef\oldfigs{\thefigs}
\global\xdef\thefigs{\oldfigs\vskip.1in\item{(\number\nfig)} #2\ }%
\fi%
\global\advance\nfig by 1 } 

\def\figurecaptions{\vfill\eject\section{Figure Captions}\thefigs}

\def\fig#1{\xdef #1{(\number\nfig)}
\global\advance\nfig by 1 } 


\newcount\ntab
\global\ntab = 1

\def\table#1{\xdef #1{\number\ntab}
\global\advance\ntab by 1 } 


\newcount\cflag
\newcount\nequation
\global\nequation = 1
\def\eqlabel{(1)}

\def\nexteqno{\ifnum\cflag = 0
\global\advance\nequation by 1
\fi
\global\cflag = 0
\xdef\eqlabel{(\number\nequation)}}

\def\lasteqno{\global\advance\nequation by -1
\xdef\eqlabel{(\number\nequation)}}

\def\label#1{\xdef #1{(\number\nequation)}
\ifnum\dflag = 1
{\escapechar = -1
\xdef\draftname{\littlefont\string#1}}
\fi}

\def\clabel#1#2{\xdef\eqlabel{(\number\nequation #2)}
\global\cflag = 1
\xdef #1{\eqlabel}
\ifnum\dflag = 1
{\escapechar = -1
\xdef\draftname{\string#1}}
\fi}

\def\cclabel#1#2{\xdef\eqlabel{#2)}
\global\cflag = 1
\xdef #1{\eqlabel}
\ifnum\dflag = 1
{\escapechar = -1
\xdef\draftname{\string#1}}
\fi}


\def\eeq{}

\def\eqnn #1\eeq{$$ #1 $$}

\def\eq #1\eeq{
\ifnum\dflag = 0
{\xdef\draftname{\ }}
\fi 
$$ #1
\eqno{\eqlabel \rlap{\ \draftname}} $$
\nexteqno}







\def\eqa #1\eeq{
\ifnum\dflag = 0
{\xdef\draftname{\ }}
\fi 
$$ \eqalignno{ #1 } $$
\global\cflag = 0}


\def\ie{{\it i.e.\/}}
\def\eg{{\it e.g.\/}}

\def\etal{{\it et.al.\/}}


\def\anp#1#2#3{{\it Ann.\ Phys. (NY)} {\bf #1} (19#2) #3}

\def\nci#1#2#3{{\it Nuovo Cimento} {\bf #1} (19#2) #3}
\def\npb#1#2#3{{\it Nucl.\ Phys.} {\bf B#1} (19#2) #3}
\def\plb#1#2#3{{\it Phys.\ Lett.} {\bf #1B} (19#2) #3}

\def\prd#1#2#3{{\it Phys.\ Rev.} {\bf D#1} (19#2) #3}

\def\prep#1#2#3{{\it Phys.\ Rep.} {\bf #1} (19#2) #3}
\def\prl#1#2#3{{\it Phys.\ Rev.\ Lett.} {\bf #1} (19#2) #3}
\def\rmp#1#2#3{{\it Rev.\ Mod.\ Phys.} {\bf #1} (19#2) #3}
\def\sjnp#1#2#3#4#5#6{{\it Yad.\ Fiz.} {\bf #1} (19#2) #3
[{\it Sov.\ J.\ Nucl.\ Phys.} {\bf #4} (19#5) #6]}


\global\nulldelimiterspace = 0pt



\def\frac#1#2{{{#1} \over {#2}}\,}  



\def\Asl{\hbox{/\kern-.7500em\it A}} 
\def\Dsl{\hbox{/\kern-.6700em\it D}} 
\def\dsl{\hbox{/\kern-.5300em$\partial$}}
\def\pxpsl{\hbox{/\kern-.5600em$p$}}
\def\sslsh{\hbox{/\kern-.5300em$s$}}
\def\epssl{\hbox{/\kern-.5100em$\epsilon$}}
\def\delsl{\hbox{/\kern-.6300em$\nabla$}}
\def\lxpsl{\hbox{/\kern-.4300em$l$}}
\def\elxpsl{\hbox{/\kern-.4500em$\ell$}}
\def\kxpsl{\hbox{/\kern-.5100em$k$}}
\def\qxpsl{\hbox{/\kern-.5000em$q$}}
\def\sla#1{\raise.15ex\hbox{$/$}\kern-.57em #1}



\def\roughly#1{\mathrel{\raise.3ex\hbox{$#1$\kern-.75em\lower1ex\hbox{$\sim$}}}}
\def\lsim{\roughly<}
\def\gsim{\roughly>}



\def\bfr{{\bf r}}



\def\Sca{{\cal A}}

\def\Scc{{\cal C}}
\def\Scd{{\cal D}}

\def\Sco{{\cal O}}


\def\ssv{{\sss V}}


\def\Tr{\mathop{\rm Tr}}



\def\Avg#1{\left\langle #1 \right\rangle}




\def\eV{{\rm \ eV}}

\def\MeV{{\rm \ MeV}}

 2
  \overfullrule=0pt


\def\bk{\item{}}

\def\Asl{\hbox{/\kern-.7500em\it A}} 

\def\GF{G_{\sss F}}

\def\dv{\delta V}
\def\dne{\delta n_e}

\def\drho{\delta\rho}
\def\spha{{Y}_{\ell m}}

\def\MSW{{\sss MSW}}
\def\VAC{{\sss VAC}}

\def\bull{\noindent $\bullet$\ }


\line{hep-ph/9707542 \hfill McGill-97/13}

\vskip .1in

\title
\centerline{Neutrino Propagation Through}
\centerline{Helioseismic Waves}
\endtitle

\vskip .2in

\authors
\centerline{P. Bamert, C.P. Burgess and D. Michaud}
\vskip .2in
\centerline{\it Physics Department, McGill University}
\centerline{\it 3600 University St., Montr\'eal, Qu\'ebec, Canada, H3A
2T8.}
\endauthors

\vskip .2in

\abstract
Motivated by earlier calculations showing large effects
when neutrinos propagate through fluctuating media, we
perform here a detailed analysis of how density
fluctuations in the sun  in the form of helioseismic waves
can modify the MSW solution to the solar neutrino problem.
We find negligible effects for the MSW spectrum, even under
extreme circumstances. There are two main reasons why
our conclusions differ from earlier analyses. First, most helioseismic
waves do not affect neutrino propagation because their amplitude
is too small in the MSW resonance region, which is the only region
to whose fluctuations neutrinos are sensitive. There is one class of
waves which may be subject to an instability, however, and so can have
significantly larger amplitudes. But the wavelength for these
waves is so long that it invalidates the previous methods of
calculation. Our more complete calculation significantly
reduces the prediction for their influence on neutrino propagation.
\endabstract


\vfill\eject

\section{Introduction and Summary}

\ref\snexps{R. Davis, D.S. Harmer and K.C. Hoffman, \prl{20}{68}{1205};\bk
J.K. Rowley \etal, in {\it Solar Neutrinos and Neutrino Astronomy},
AIP Conference Proceedings number 126, edited by M.L. Cherry, W.A. Fowler
and K. Lande, (1985); \bk
K.S. Hirata \etal, \prl{65}{90}{1297}; \bk
P. Anselmann \etal, \plb{327}{94}{377}; \bk
J.N. Abdurashitov \etal, \plb{328}{94}{234}.}

\ref\SNTheory{The following references provide excellent reviews: \bk
J.N. Bahcall, {\it Neutrino Astrophysics}, Cambridge University Press, 1989;
\bk
S. Turck-Chi\`eze \etal, \prep{230}{93}{57};\bk
W. Haxton, {\it Ann. Rev. of Astr. and Astrophys.} {\bf 33} (1995) 459.}

Taken together, the combined information from the four pioneering
solar neutrino experiments \snexps\ suggests a neutrino spectrum which
deviates strongly from astrophysical predictions \SNTheory. Purely
astrophysical
attempts to reconcile the data with theory fail because of the virtually
complete suppression of ${}^7$Be neutrinos seen by the experiments together
with the observed nonzero ${}^8$B neutrino flux. These two observations
are hard to reconcile with one another using a solar model since the observed
${}^8$B neutrino flux is made from the same chain of nuclear reactions
which would have also produced the missing ${}^7$Be.

\ref\msw{L. Wolfenstein, \prd{17}{78}{2369};\bk
V. Barger, K. Whisnant, S. Pakvasa and R.J.N. Phillips, \prd{22}{80}{2718};\bk
P. Langacker, J.P. Leville and J. Sheiman, \prd{27}{83}{1228};\bk
S.P. Mikheyev and A. Yu. Smirnov, {\it Sov. Phys. Usp.} {\bf 29} (1986) 1155;
\sjnp{42}{85}{1441}{42}{85}{913};
\nci{9}{86}{17};\bk
S.P. Rosen and J.M. Gelb, \prd{34}{86}{969};\bk
H. Bethe, \prl{56}{86}{1305};\bk
W. Haxton, \prl{57}{86}{1271};\bk
A.J. Baltz and J. Weneser, \prd{37}{88}{3364};\bk
P.D. Mannheim, \prd{37}{88}{1935}.}

Particle physics explanations of the neutrino spectrum don't suffer
from the same difficulty. The most successful of these explanations
is the resonant conversion of neutrino flavours within the solar medium,
the so called MSW mechanism \msw . It provides an excellent description
of the experiments using theoretically plausible neutrino parameters.

\ref\noisy{F.N. Loreti and A.B. Balantekin, \prd{50}{94}{4762}
(nucl-th/9406003);\bk
F.N. Loreti, Y.Z. Qian, G.M. Fuller and A.B. Balantekin, \prd{52}{95}{6664}
(astro-ph/9508106);\bk
E. Torrente Lujan, preprint BUTP-96-8 (hep-ph/9602398).}

\ref\denis{D. Michaud, McGill University M.Sc. thesis, 1994;\bk
in the 17th proceedings of the {\it Annual MRST Meeting}, 1995.}

\ref\noisytwo{H. Nunokawa, A. Rossi, V.B. Semikoz and J.W.F. Valle,
\npb{472}{96}{495} (hep-ph/9602307).}

\ref\noisythree{A.B. Balantekin, J.M. Fetter and F.N. Loreti,
\prd{54}{96}{3941} (astro-ph/9604061).}

\ref\cliffdenis{C. P. Burgess and D. Michaud, \anp{256}{97}{1},
(hep-ph/9606295);\bk
and contribution to the proceedings of the 17th International Conference on
Neutrino Physics and Astrophysics (NEUTRINO '96), ed.
by K. Enqvist, K. Huitu and J. Maalampi, Helsinki, World Scientific,
(hep-ph/9611368), 1996.}

The original MSW analysis starts from a mean-field treatment of
the background through which the neutrinos propagate. More recent
work has since investigated how corrections to the mean-field
picture might influence the neutrino survival probability
\noisy, \denis, \noisytwo, \noisythree, \cliffdenis. For
technical reasons these studies focus on fluctuations about
the mean which have small spatial correlation lengths
compared to the distances over which neutrinos appreciably
oscillate. Within this approximation the conclusions obtained
were that fluctuations of this type in the solar electron density
can indeed significantly modify the MSW neutrino spectrum,
provided their amplitude near the MSW resonance point can
be as large as a few percent of the background density.

\ref\czturbulence{See, \eg, H.C. Spruit, {\it Mem. S.A. It.} {\bf }
(1996), astro-ph/9605020.}

\ref\bahcall{J.N. Bahcall, M.H. Pinsonneault, S. Basu and
J. Christensen-Dalsgaard, \prl{78}{97}{171}.}

The missing element to these studies has been the identification
of a realistic source of such fluctuations within the sun. Unfortunately,
two things work against having fluctuations affect neutrino
propagation in the sun. First, ordinary thermal fluctuations give
negligibly small contributions \cliffdenis\ because of the extremely
small interaction cross section between neutrinos and other
particles at solar densities. Second, in order to appreciably
disturb neutrino oscillations,
macroscopic fluctuations would have to occur in the
same parts of the sun as does the MSW resonance itself
\cliffdenis. Unfortunately,
most of the fluctuation sources which have been proposed
are associated with the enormous turbulence of the sun's
outlying convective layer \czturbulence, and so
are too far from the MSW resonance regions to have much effect.
More exotic proposals which involve
mixing and convection in the solar core, have recently
been ruled out by helioseismic data \bahcall.

\ref\solwaves{A very readable review is:
Jorgen Christensen-Dalsgaard, Lecture Notes,
available at \bk {\tt http://bigcat.obs.aau.dk/$\sim$jcd/oscilnotes/.}}

To our knowledge, there is only one source of fluctuations
which is known to be present in our sun, and which is not
known to be excluded from the solar interior: helioseismic waves
\solwaves. Furthermore, estimates based on the short-correlation-length
analysis indicate \noisytwo, \cliffdenis\ that it would suffice
for such waves to have amplitudes of a few percent in order for
them to alter solar-neutrino propagation.

The present paper provides a more careful calculation of
solar-neutrino propagation through realistic helioseismic
waves, to accurately determine the potential size of
their influence on MSW oscillations.  We explicitly compute
seismic waves profiles within the sun in order to reliably
correlate their amplitude at the resonance point with
observations (which are made at the solar surface). We
then numerically evolve neutrinos through these
profiles to identify their effect on MSW oscillations.
Since it turns out that most of the waves of interest have
wavelengths much longer than typical neutrino oscillation scales,
we perform the calculation without making use of the
short-correlation-length approximation.

Our results, in a nutshell, are as follows:

\ref\parres{P. Krastev and A.Yu. Smirnov,
\plb{226}{89}{341};\bk
W. Haxton and W.-M. Zhang, \prd{43}{91}{2484}.}

\bull\ Standard MSW neutrino oscillations are {\it not}
affected by helioseismic waves,\foot\parres{One way
small effects can accrue incrementally to appreciably
influence neutrino oscillation is if they act repeatedly,
such as for a parametric resonance \parres. This
kind of mechanism appears to play no role when neutrinos
pass through helioseismic waves having realistic
amplitudes, however.} for the following reasons.

\bull\ The observed helioseismic waves ($p$-waves)
{\it decrease} in amplitude as one moves into the sun,
with the result that their observed size at the solar
surface makes them too small to produce observable
effects in the resonance region.

\bull\ An important class of waves ($g$-waves) {\it grows}
in amplitude with increasing solar depth.
Furthermore, some of these
may be subject to an instability which makes their amplitude
orders of magnitude larger than their stable compatriots.
The wavelengths of these waves are typically too large
to permit working in the small-correlation-length limit.
However, it turns out that small-correlation-length
calculations, when applied beyond their domain of
applicability, give an {\it overestimate} of neutrino-propagation
effects, and so a full calculation gives negligible effects
for neutrinos, even from these strongly-excited modes.

\bull\ Furthermore we find that a density perturbation most strongly
affects neutrino propagation when its correlation length is of the order
of the length scale at which the short-correlation-length approximation fails,
typically within an order of magnitude of the neutrino oscillation length.

The outline of the rest of this paper is as follows. In \S2
we recap how density fluctuations can influence neutrino
propagation, including a comparison between `exact' results
and those obtained using the formalism adapted to short
correlation lengths (called henceforth the
{\it master-equation} approach). This comparison shows
precisely how long correlation lengths must be to invalidate
the master-equation approach, as well as how bad its
predictions become when applied beyond its domain of
applicability. \S3 gives a whirlwind summary of what is
known about the internal structure of the sun, and the
seismic waves it supports. \S4 describes the numerical
aspects of our simulations of neutrino propagation
through helioseismic waves.  The resulting conclusions are
then drawn in \S5.

\section{Neutrino Propagation in the Presence of Density
Fluctuations}

This section gives a short review of the influence
of density fluctuations on solar neutrino propagation. We
treat separately the case where the correlation length is
short, and describe the master-equation formalism which
applies in this limit. The results in this limit are then compared
with a more complete calculation for a simple model of
solar fluctuations.

\subsection{Solar Neutrinos and Fluctuations}

Conceptually, there are two distinct types of fluctuations
to consider when any particle propagates through a medium
like the sun. First, if the particle frequently interacts
with the medium's constituents, and these constituents
are randomly distributed, then the effects on {\it a
specific particle} of multiple interactions with the
medium may be described by a statistical average
over a fictitious ensemble of equivalent media. An
example of this sort is furnished by photons propagating
through the sun, since each photon scatters many
times while escaping from the solar interior. Although
this kind of multiple scattering is believed to happen
to neutrinos moving through the interior of a supernova,
it does not apply to neutrinos in the sun.

Alternatively, if the mean properties of the medium
vary in space or time, the mean features as seen
by successive particles can vary. The average
response of a detector to many such
particles may also be described in terms of an
average over the ensemble of `media' seen
by the individual particles.
This is the kind of average which is relevant for solar
neutrino propagation through helioseismic waves.
Any one neutrino interacts extremely rarely, and
escapes the sun in a matter of seconds.
Helioseismic waves, on the other hand, alter the
electron density over timescales of minutes or hours,
and vary in space over distances comparable with
the solar radius.

Since any one neutrino interacts so weakly with
the solar environment as to
see negligible fluctuations, the time evolution of its
density matrix is described by the usual Schr\"odinger
equation:
\label\Schreqn
\eq
\rho(t,n) = U(t,n) \, \rho(0) \, U^\dagger (t,n),
\eeq
where $\rho$ is the neutrino density matrix, $t$ denotes
time, $U(t,n)$ is a unitary evolution operator, and $n$
generically denotes all of those features of the solar
medium on which the evolution can depend parametrically,
and which can change in the time between the transit of
different neutrinos. For example, for neutrino
oscillation experiments $\rho$ may be taken to be
a matrix in neutrino-flavour space, and $U$ is the
solution to the evolution equation
\label\Ueveqn
\eq
{\partial U \over \partial t} = -i V(t,n) \, U
\eeq
where $V(t,n) =  V_\VAC + V_\MSW(t,n)$, with
$V_\VAC \approx k+{m^\dagger m\over 2k}+\dots$ giving the
vacuum evolution matrix for an ultrarelativistic neutrino
of three-momentum $k$ and mass matrix $m$. Similarly,
$V_\MSW(t,n) \equiv \sqrt{2}\,\GF g^e\Avg{n_e(t)}$ is the
leading-order effective interaction with the mean matter
background, and $g^e={\rm diag}(1,0)$ is a matrix describing
the charged-current coupling to the electron density
$n_e(t)$. For simplicity we consider here only
the case of two active neutrino species.

The final response of a neutrino
detector is then obtained by averaging the
appropriate observable over an ensemble of values
for the variables, $n$, weighted by a probability
distribution, $p(n)$:
\label\obsavg
\eq
\Avg{\Sco}(t) = \int dn \; p(n) \; \Tr \Bigl[ \Sco
\;  \rho(t,n) \Bigr] = \Tr \Bigl[ \Sco
\;  \Avg{\rho(t)} \Bigr] .
\eeq

Much of the physics of the resulting fluctuations is encoded in
the probability distribution, $p(n)$. The most commonly-used
choice \noisy, \noisytwo, \noisythree, \cliffdenis\ is to
assume density fluctuations which are uncorrelated in space.
For the present applications we instead follow
Ref.~\cliffdenis\ and expand the electron density in terms
of a complete basis of helioseismic modes:
\label\basis
\eq
n_e(t) = \Avg{n_e(t)} \left[1+\sum_j \Scc_j \phi_j(t) \right] ,
\eeq
where the coefficients $\Scc_j$, are assumed to be
uncorrelated random variables which are Gaussian distributed,
with vanishing mean:  \ie\ $\Avg{\Scc_j}=0$  and
$\Avg{\Scc_j\Scc_k}=\Scd_j\delta_{jk}$. \S3 is devoted to
the explicit construction of the basis functions which
are appropriate for helioseismic waves.

In subsequent sections we directly solve these equations for
$U(t,n)$ and $\rho(t,n)$ to find the electron-neutrino survival
probability after transit through the sun.  Before doing so,
we next pause to describe the master-equation approach,
which has been the main theoretical tool used in previous
investigations.

\subsection{Short Correlation Lengths: The Master Equation}

We summarize here the formalism of \noisy,
\noisytwo, \noisythree\ and \cliffdenis,  in which references
the interested reader may find more details.

The master-equation approach arises most naturally for
physical systems where the ensemble average is meant
to describe a transient particle's multiple interactions
with the ambient material. For this kind of system
it is natural to formulate a a {\it coarse-grained}
time derivative, $D \rho/Dt$, of the density matrix
which incorporates the average over short-timescale
fluctuations. The result turns out to give a reasonable
description of solar neutrino propagation through
short-range density variations, even though neutrinos do
not multiply interact with the solar medium.

The starting point is the observation that treating the local
electron density as a random variable introduces a correlation
length into any ensemble averages. More precisely, suppose
that the correlation $\Avg{\dv(t) \, \dv(t')}$ (where
$\dv = V - \Avg{V}$) becomes negligible whenever
$|t - t'|$ is greater than some characteristic scale,
$\tau_c$. When coarse-grained over times longer
than $\tau_c$, the derivative, $D \rho/Dt$, evaluated
at time $t$, depends only on the value of $\rho(t)$, and
has no memory of how $\rho$ has evolved over
earlier times. Furthermore, when $\tau_c$ is sufficiently
short compared to the time scales in the interactions,
the coarse-grained derivative,  $D \rho/Dt$, may
also be evaluated perturbatively in $V$.

As applied
to the neutrino density matrix in two-by-two
flavour-space, this leads to the {\it master equation} \noisy,
\noisytwo, \noisythree, \cliffdenis:
\eq
\label\drhoflavour
{D \rho \over Dt} =-i \Bigl[ V_\VAC + V_\MSW(t) ,\rho \Bigr]
-2 \, \GF^2 \Sca(t) \Bigl[ (g^e)^2\rho + \rho (g^e)^2 - 2 \, g^e
\rho g^e \Bigr] + {\cal O}(V^3).
\eeq
The first term of eq.~\drhoflavour\ describes the usual
MSW evolution.
The coefficient $\Sca(t)$ of the
second term is the correlation integral
\eq
\label\correlation
\Sca(t)\equiv\int_{t'}^t  d\tau \; \Avg{\dne(t)\dne(\tau)}
\eeq
which represents the fluctuation effects.

\ref\Parkeref{S.J. Parke, \prl{57}{86}{1275}.}

Eq.~\drhoflavour\ can be directly integrated if the following two
approximations hold: ($i$) the neutrino evolution away from MSW
resonances can be treated in the adiabatic approximation, and ($ii$)
the resonance region is sufficiently narrow as to justify approximating
the electron density profile there as a linear function of distance
along the neutrino path.  In this
limit one obtains a prediction for the probability that an
electron-neutrino produced at time $t'$, survives to time $t$,
which generalizes the well-known Parke formula
\Parkeref\ to include the effects of fluctuations \cliffdenis:
\eq
\label\parke
P_e(t,t') = {1\over 2} + \left( {1\over 2} - P_J \right) \lambda
\cos 2\theta_m(t')\cos 2\theta_m(t) .
\eeq
Here $P_J = \exp\left[ -{\pi\over 2} \left({\sin^22\theta_\ssv \over
\cos 2\theta_\ssv} \right) \left( {\delta m^2 h \over 2k} \right) \right]$
is the jump probability, where $h$ is the scale height for the
averaged electron density, and $\delta m^2$ is the squared-mass
difference between the two neutrino mass eigenstates in vacuum.
$\theta_\ssv$ is the vacuum mixing angle, while $\theta_m(t)$
is the matter mixing angle evaluated at the position occupied
by the neutrino at time $t$. The coefficient,
$\lambda$, equals one in the absence
of fluctuations, but more generally is given by
\eq
\label\lamb
\lambda\equiv \exp\left[ -2 \, \GF^2 \int_{t'}^t  d\tau \;
\Sca(\tau) \sin^22 \theta_m(\tau) \right] .
\eeq
Notice that eq.~\lamb\
implies that fluctuation effects are strongest
at resonance where $\sin^22\theta_m$ is maximised.
We find in subsequent sections that this conclusion
still holds when we go beyond the
limitations of this perturbative approach.

\subsection{When the Master-Equation Technique Fails}

Eq.~\drhoflavour, and so also eqs.~\parke\ and \lamb,
fail once the correlation time, $\tau_c$, becomes
too large. A simple way to see that this must be so
is to notice that $\Sca(t)$ need not, in general,
be positive, and so
for large correlation lengths --- and so for large
$\Sca(t)$ --- $\lambda$ can be greater than one. This,
in turn, can make $P_e$ lie outside the interval
$[0,1]$, and so be unphysical. This section aims to
determine more precisely when, and by how much,
predictions based on the master equation, eq.~\drhoflavour,
fail once correlation lengths become large.

A first estimate for when these formulae fail
may be obtained by figuring the size of the $O(V^3)$
term in $D\rho/Dt$. For the purpose of so doing,
we consider a simple model of fluctuations in
which the solar medium is divided into regions
(`cells') of linear size $\ell$. We imagine an
ensemble of density fluctuations in these
cells, which are uncorrelated from cell to cell.
Quantitatively, we take
\label\celldef
\eq
\Avg{\dne(\bfr)\dne(\bfr')} = \left\{
\matrix{ \epsilon^2 \Avg{n_e(\bfr)}
\Avg{n_e(\bfr')}  & \hbox{if $\bfr$ and $\bfr'$ lie
in the same cell;} \cr 0 & \hbox{otherwise} \cr}
\right.   .
\eeq
With this choice, the fluctuation term in eq.~\drhoflavour\ is
of order $\GF^2\Sca(t) \sim \ell (\epsilon\GF \Avg{n_e})^2$, while
the neglected $O(V^3)$ terms are $\sim \ell^2 (\epsilon\GF
\Avg{n_e})^3$.

This very crude estimate therefore indicates that
eq.~\drhoflavour\ fails for fluctuations satisfying
$\ell \gsim 1/(\epsilon\GF \Avg{n_e})$. This estimate
agrees passably well with the more detailed
comparison between the master-equation method,
and an `exact' ensemble average, which we now
present.

To obtain a feel for the numbers, taking
$1/(\GF \Avg{n_e}) \sim 300{\rm km}$, which is
typical at resonance, and $\epsilon \sim 10 \%$
this condition becomes
$\ell > 3,000 {\rm km}$. Helioseismic waves,
especially those with the largest amplitudes,
have wavelengths (\ie\ correlation lengths)
at resonance which can be of order $R_\odot/10 \sim
70,000$ km, and so easily exceed the size of $\ell$.

\figure\figcomp{Here we compare the two averaging
procedures described in the text in the context of
the 'Cell' model. We show the survival probability as a
function of cell size (\ie\ correlation length), $\ell$,
using the following set of neutrino parameters:
${\delta m^2 / 2k} \sim 10^{-6} \; \eV^2/\MeV$,
$\sin^2 2\theta_\ssv =0.01$ and $\epsilon = 0.1$.
$\Avg{n_e}$ was chosen to fall exponentially along
the neutrino-propagation ($z$) direction, with
scale height $h = R_\odot/10 \sim 6.6 \times 10^4$
km.
The horizontal dotted line gives the MSW result,
while the result obtained using the generalized Parke
formula is given as a thin solid line. The other
three curves give the result obtained using the ordinary
Parke formula for each neutrino, with the result
averaged over a random ensemble of
200 density profiles of the Cell type.
The thick solid line adjusts the
cells to ensure that the MSW resonance always
occurs inside a cell. The dashed curve similarly
adjusts the cells to ensure that the resonance
occurs precisely at a cell boundary. The dot-dashed
curve gives the result for randomly distributed
cell positions. The differences between these
curves show the sensitivity of the result to
the details of how the cells are defined, and in
particular to the sharp density gradient we
have assumed between cells.
The generalized Parke formula differs significantly
from the others for cell sizes larger than about
$10^4$ km. Above this point the generalized Parke
value for the survival probability remains fixed
around $P_e=0.5$, whereas the 'real', numerically
obtained, value decreases sharply to the MSW result.
Notice also that a large density gradient right
at the resonance, given here by a sharp cell
boundary, can have a significant effect even for very
large correlation length. }

We have performed a more quantitative comparison
between neutrino-survival predictions using the
master equation, or a direct numerical ensemble
average performed without approximation
over 200 random density profiles of the
Cell type. We choose a geometry,
in which the neutrino moves along the $z$ axis
through a grid of rectangular cells of length $\ell$.
We take the mean density profile to fall exponentially
with $z$, with a scale height typical of solar models:
$h \sim 6.6 \times 10^4 {\rm\ km}\sim R_\odot/10$. The
fluctuation size is taken as $\epsilon = 0.1$.

Figure \figcomp\ summarizes the results of this comparison.
The horizontal dotted line in this figure gives the standard MSW survival
probability in the absence of fluctuations, using the oscillation
parameters ${\delta m^2/2k} \sim 10^{-6}
\; \eV^2/\MeV$ and $\sin^2 2\theta_\ssv =0.01$. The thin solid
line expresses the prediction of the generalized Parke
formula, eqs.~\parke\ and \lamb. Notice how the
generalized Parke expression approaches the MSW
result for small $\ell$, and goes to a fixed value, $P_e = 0.5$,
for large $\ell$.

For comparison, the remaining three curves give the results
of a direct numerical average over an ensemble of electron
densities, without using the master-equation formalism.
For these curves the survival probability for each neutrino
is computed using Parke's formula --- in the form
originally proposed by Parke, without fluctuations ---  and
this result is then numerically averaged
over 200 elements of the density ensemble. We have
checked our use of the ordinary Parke formula for the
survival probability of any one neutrino as it moves
through a given density profile by comparing it with
the direct numerical integration of the neutrino
evolution equations. We find good agreement, although
for large-amplitude helioseismic waves
this agreement can require the use of Parke's formula
for the passage through several MSW resonances.

The three curves obtained in this way
differ in how they treat the case where the MSW
resonance falls at the boundary of a cell, where the density
profile is varying artificially strongly. In the thick solid
line, the cells are adjusted to ensure that the MSW resonance
always falls well away from any cell boundary. By contrast,
the dashed line arranges the resonance to always fall
at the cell boundary. The dot-dashed line corresponds to
randomly positioned cells, whose boundaries may or may
not fall at the MSW resonance. Notice these three curves
agree with each other except for the largest $\ell$,
where the larger effect is obtained when the resonance
hits a cell wall. This cell-boundary effect
is less pronounced for smoother density profiles.
In the absence of cell wall/resonance
coincidences (thick solid curve) the ensemble-averaged
survival probability approaches the MSW prediction for
sufficiently large $\ell$.

The generalized Parke prediction differs most dramatically from
the direct ensemble average for $\ell \gsim 5,000$ km,
as may be seen by comparing the thick and thin solid lines
in Fig.~\figcomp. For larger $\ell$ the numerical simulation
agrees with the MSW prediction (in the absence of cell-boundary
effects); a feature which is missed by the generalized Parke result
(thin solid line).
Notice that 5,000 km agrees reasonably well with the estimate,
$\ell \gsim 1/(\epsilon \GF \Avg{n_e}) \sim 3,000$ km, given above,
for the scale at which correlations should fail to be well described
by master-equation methods.

The lessons from this section are these:

\bull\ Master equation methods describe solar
neutrinos surprisingly well for sufficiently small correlation
lengths, but can dramatically overestimate the deviation
from the MSW prediction when the correlation lengths
are large.

\bull\ Numerically, the dividing line between large and
small correlation lengths is of order 3,000 km, which
is quite small compared to the length scales over which
helioseismic waves can vary.

\section{Helioseismic Waves at a Glance}

\figure\figreson{This figure indicates the position of the
MSW resonance within the sun as a function of neutrino energy
and MSW parameters. The two solid lines delimit the range
permitted for the small-angle MSW solution, whereas the
dashed and dotted lines represent various options within
the large-angle MSW solution. The main production regions for
$p$-$p$, ${}^7$Be, and ${}^8$B neutrinos are indicated for
comparison in the same figure as shaded areas.}

\ref\solmod{S. Turck-Chi\`eze \etal, \prep{230}{93}{57}.}

Our sun is not as quiet as it seems. On the contrary, it displays
a wide variety of nonequilibrium phenomena, many of which
might give rise to density perturbations \solmod .
Unfortunately, most of these solar disturbances are
restricted to the solar surface, or to the convective
zone (\ie\ $r/R_\odot \gsim 0.7$). By contrast, Figure
\figreson\ shows the depths of the MSW resonance
regions as a function of neutrino energy and mixing
parameters, illustrating that all resonances occur
well within half a solar radius. It follows that most
solar density fluctuations are much too close
to the solar surface to play any role
in neutrino oscillations.
The notable exception to this are helioseismic waves, the
enumeration of whose properties is the main subject of this section.

\subsection{Qualitative Discussion}

We next present the main features of helioseismic waves,
with our treatment following the excellent review found
in Ref.~\solwaves.

We start with a brief reminder of the solar structure. The sun
can be divided into four main regions. The energy-producing core
is at the centre, extending out to roughly $0.2$ solar radii.
Next comes the radiative zone, for $0.2 \lsim r/R_\odot \lsim
0,7$, in which the dominant means of energy transfer is
radiative. The convective zone follows, for $r/R_\odot \gsim
0.7$, where convection is the dominant method of energy
transfer. Finally, the solar surface occupies a comparatively
thin layer near $r/R_\odot \sim 1$. The enumeration of these
regions is important because the properties of a helioseismic
wave are largely governed by the zone in which it sits.

Solar oscillations can be classified into two qualitatively
different types of waves, known as `$p$-' and `$g$-waves'
respectively, depending on the nature of the restoring force
which is responsible for the oscillatory behaviour. For
$p$-waves pressure is the relevant restoring force,
while for $g$-waves it is buoyancy which plays this role.
Because of this difference in restoring force for each, $p$-
and $g$-waves tend to be found in different regions of the
sun. Since convection is related to an instability in the
buoyancy force, $g$-waves are damped within the convective
zone, and so are confined to the interior of the sun.
$p$-waves, on the other hand, can exist in all four zones,
although because the speed of sound increases with depth,
only radially-directed waves tend to penetrate deep
into the central regions. Both have similar frequencies,
with $p$-waves having periods typically below 30 minutes,
and $g$-waves with periods longer than this value.

\subsection{Quantitative Features}

Modelling the properties of these waves requires solving
the four hydrodynamic equations --- expressing the
conservation laws of particle number, energy and
momentum ---  which govern their properties.
The approximate spherical symmetry of the sun
allows the solutions to be expressed in terms of
spherical harmonics so long as only small perturbations about
the static background are considered.
The density perturbation, $\delta \rho$, representing a
wave can thus be written
\eq
\label\waves
\drho(r,\theta,\phi,t) = \sqrt{4\pi} \; \varrho_{n\ell}(r)
\spha(\theta,\phi) \exp(-i\omega_{n\ell} t) ,
\eeq
where the spherical harmonics $\spha$ are normalized to
satisfy $\int d\phi \, d\theta \, \sin \theta \, |\spha|^2 = 1$.
Spherical symmetry ensures that the wave frequency,
$\nu = \omega/(2\pi)$, depends only on the radial and
angular quantum numbers, $n$ and $\ell$, but not the azimuthal
quantum number, $m$. This degeneracy of modes
is broken by the rotation of the sun, although this effect
is too small to be of further relevance here.

Our simulations are performed using the Cowling approximation,
in which the gravitational backreaction of the wave onto itself
is neglected. In this approximation, the hydrodynamical
equations simplify considerably when they are expressed
in terms of the following variable: $\xi={ |S_\ell^2/ \omega^2-1|
\over \rho_0 c^2 r^2} \;\xi_r$, with $\xi_r$ denoting the
radial displacement of the wave, $\rho_0$ representing the
unperturbed mass density and $c$ denoting the speed of
sound of the background solar medium. The equations can then be
written \solwaves:
\eq
\label\cowling
\xi''(r) + K(r) \xi = 0, {\rm\ \ with \ \ }
K(r) \approx {\omega^2 \over c^2(r)} \left(
{N^2(r) \over \omega^2}-1 \right) \left({S^2_\ell(r)+
\omega_c^2(r) \over \omega^2}-1 \right) .
\eeq
The buoyancy frequency, $N(r)$, and acoustic frequency,
$S_\ell(r)$, vary with depth within the sun, and are given
explicitly by the following expressions:
\label\freqdefs
\eq
N^2 = {g (P_0' - c^2 \rho_0') \over \rho_0 c^2 } ,
\qquad \hbox{and} \qquad
S^2_\ell = {\ell (\ell+1) c^2 \over r^2} .
\eeq
In these expressions $'$ denotes differentiation
with respect to $r$, $P_0(r)$ is the background
pressure, and $g$ is the local acceleration due to gravity.
$\omega_c$ is an acoustic cutoff frequency which enters
when the waves are not permitted to propagate beyond the
solar surface. It therefore only plays a minor role
deep inside the sun. Notice that $N^2 < S_\ell^2 +
\omega_c^2$ is satisfied everywhere within the sun.

Once the radial displacement, $\xi_r$, is obtained by solving
eq.~\cowling, the radial part of the density perturbation,
$\varrho(r)$, is found (except for very small $r$)
using the following relation:
\eq
\label\rhor
\varrho(r) = {1\over (S_\ell^2-\omega^2)r^2} \left[
\omega^2 {d\over dr} - {S_\ell^2N^2 \rho_0 \over dP_0/dr}
\right] \left(r^2\rho_0 \xi_r\right) .
\eeq

\figure\figsolfreq{The solid line gives the buoyancy
frequency, $N(r)/2\pi$, as a function of depth within the sun.
The dashed lines similarly plot the acoustic frequency,
$S_\ell(r)/2\pi$, for different values of $\ell$($\equiv$L).
The two horizontal, dot-dashed lines represent the
radial interval over which representative $p$- or $g$-waves
would be undamped. The $p$-wave indicated was assumed
to have angular degree $\ell = 10$.}

Eq.~\cowling\ is easily solved in the WKBJ approximation,
with solutions
\label\wkbjsolns
\eq
\xi^\pm (r) = K(r)^{-1/4}\exp \left[ \pm i\int^r
dx \, \sqrt{K(x)} \right] .
\eeq
In this approximation the solutions therefore propagate
when $K>0$, and are damped when $K<0$. This implies the
existence of two separate frequency intervals
for propagating waves, depending on the size of $\omega$
in comparison with $N$ and $S_\ell$:
\label\pgmodes
\eq
\eqalign{ \omega<|N|\qquad\; \qquad &: \qquad\hbox{$g$-waves} \cr
\omega > \sqrt{S_\ell^2 + \omega_c^2} \qquad &: \qquad
\hbox{$p$-waves} \cr}
\eeq
The turning points for each kind of wave (which delimit
the propagating regions) are defined as those radii
for which $K$ changes sign, \ie\ those $r$ for which
$\omega^2 = N^2(r)$ or $\omega^2 = S_\ell^2(r) + \omega_c^2(r)$.
Figure~\figsolfreq\ plots the frequencies $N(r)$ and $S_\ell(r)$
as functions of radius within the sun, from which information
the positions of the turning points for the various helioseismic
waves may be inferred.

Because they are confined within the solar volume, helioseismic
waves have quantized frequencies. Within the WKBJ
approximation the allowed frequencies are determined by the
condition:
\eq
\label\duvall
\int_{r_1}^{r_2} K(r)^{1/2}dr = (n-{1\over 2}) \;\pi;
\qquad n=1,2,\dots \, .
\eeq
Here $r_{1,2}$ denote the wave's turning points, and $n$ is its
radial quantum number.

Typically, the larger the radial and angular quantum numbers,
$n$ and $\ell$, are, the better the various approximations
we have used --- such as Cowling and WKBJ --- become.
We emphasize, however, that for the purposes of the
present paper, all of the approximations used above are sufficiently
accurate even for small $\ell$ and $n$.

\figure\figmodesp{This figure plots the amplitude of
a few sample $p$-waves against radius in the sun.
The vertical axis gives the scaled quantity $\xi_r(r) r
\rho_0^{1/2}(r)$ in arbitrary units, where $\xi_r(r)$ is
the wave's radial displacement and $\rho_0(r)$ is the
background mass density.}

\figure\figmodesg{This figure plots the amplitude of
a few sample $g$-waves against radius in the sun.
The vertical axis gives the scaled quantity $\xi_r(r) r
\rho_0^{1/2}(r)$ in arbitrary units, where $\xi_r(r)$ is
the wave's radial displacement and $\rho_0(r)$ is the
background mass density.}

We obtain numerical solutions of eq.~\cowling\
by matching linear combinations of the solutions,
eq.~\wkbjsolns, across the  turning points.
Figure \figmodesp\  plots some sample $p$-wave profiles
which were found in this way. Figure \figmodesg\
gives similar plots for a few sample $g$-waves.

Several general properties of both $p$- and $g$-waves
emerge from an inspection of Fig.~\figsolfreq, and
from the above discussion:
\item{1)}
The square of the buoyancy frequency, $N^2$, becomes
negative inside the convective zone. Indeed, the instability
which this implies for the corresponding $g$-modes
is precisely the instability towards convection which
defines the convective zone. It is a general feature,
then, that $g$-waves are damped inside the convective
zone, and so are much harder to observe on the solar
surface than are $p$-waves. This is the main reason
why $p$-waves have been observed on the solar surface
while $g$-waves have not.
\item{2)}
The maximal value obtained by $N(r)$ is $N_{\rm max}/2\pi
\sim 500 \; \mu{\rm Hz}$, which is the largest frequency
possible for $g$-waves. Similarily, since
$S_\ell^2 + \omega_c^2 > N^2$,
$N_{\rm max}$ also gives a lower bound on possible
$p$-wave frequencies.
\item{3)}
Since $S_\ell$ increases with $\ell$ and decreases with $r$
only $p$-waves having small $\ell$ can penetrate very
deeply into the solar interior.
\item{4)}
Since $g$-waves also have frequencies which
grow with $\ell$, those with large $\ell$ typically have
frequencies which saturate the allowed maximal value,
$\omega \sim N_{\rm max}$. As a result, we also expect
these waves to have their largest amplitudes for radii in
the interval $0.1 < r/R_\odot < 0.3$. This makes these
waves potentially interesting for neutrino propagation
since this is precisely where MSW resonance typically occurs.
Furthermore, these waves are also damped more strongly
in the convective zone since $\varrho \sim r^{-(\ell+3/2)}$,
and so they are even harder to detect
using measurements at the solar surface.

\subsection{Amplitudes at Resonance and Relevance for
Neutrino Propagation}

We are now in a position to answer the key question.
How big can these waves be near the neutrino resonance
point? The answer to this question dictates the size
of the contribution of these waves in the detailed
simulations of the next section.  The answer to this
question is necessarily different for $p$- and $g$-waves,
since experimental observations are available for
$p$-waves, but not for $g$-waves. We therefore
handle each of these waves separately in what
follows.

\topic{Pressure waves}

\ref\soho{B. Fleck, {\it Rev. Mod. Astron. }{\bf 10} (1997) in press,
{\tt http://esa.nascom.nasa.gov /$\sim$bfleck/Preprints/ }.}

\figure\figpwaves{Here we plot the maximum amplitude which is
attained by a pressure wave for $r < 0.5 \, R_\odot$. The waves
are normalized by requiring them to reproduce the
experimentally-measured energy distribution
as a function of frequency (see \eg\ \solmod).
The interval chosen for $r$
is sufficiently generous to contain the MSW resonance for virtually
all choices for neutrino mixing parameters. The points displayed
give $\drho/\rho$ as a function of linear frequency,
$\nu_{n\ell} = \omega_{n\ell}/2\pi$, for $p$-waves
having various angular ($\ell\equiv$L) and radial ($n$) quantum numbers.
To guide the eye, solid lines connect points representing waves
having a fixed value of $\ell$($\equiv$L), but varying $n$. }

Pressure waves are undamped in the convective zone, and so
propagate right up to the surface where they can be observed.
A huge number of these modes have indeed
been identified by a number of different experiments
(see \eg\ \soho, as well as the references found in \solmod). The
measured surface velocities of individual modes are found
to be in the range of $10$ cm/sec.

\ref\enerprof{K.G. Libbrecht, M.F. Woodard and J.M. Kaufman,
{\it Astrophys. J. Suppl.}{\bf\ 74}, (1990) 1129.}

We determine the amplitudes
of these waves by requiring that their energy reproduces the
observed energy distribution as a function of frequency~\enerprof\
(see also~\solmod ).
This distribution is observed to peak at a value
$10^{28}$ erg when $\nu \sim 3$ mHz.

We plot the resulting amplitudes in Fig.~\figpwaves. Here
we display the maximal amplitude which a $p$-wave can
obtain anywhere within the innermost $0.5$ solar radii, given
that it is normalized to reproduce the observed energy
distribution. Each dot on this figure gives the frequency
and the maximum
amplitude of a wave having a specific value of $n$ and $\ell$.
Lines connecting dots which differ only in $n$, but not in
$\ell$, are drawn to help guide the eye.

As may be clearly seen from this figure, the maximum
wave amplitudes are typically of order of $\drho/\rho_0 \lsim
10^{-10}$, making these waves irrelevant for neutrino propagation.

\topic{Buoyancy waves}

\ref\gmod{P. Kumar, E.J. Quataert and J.N. Bahcall,
{\it Astrophysical Journal} {\bf 458}, (1996), L83.}

Since buoyancy waves have not been observed, a determination
of their amplitude is necessarily more uncertain. In this
section we estimate their potential amplitude in two steps.
First we ask how big it is possible for them to be without
their having been detected. Then we ask how big they
might plausibly be expected to be on theoretical
grounds.

For the purposes of determining how large a $g$-wave could
escape detection, we demand that the radial velocity
of oscillation at the surface of the sun produced by the wave
be less than 1 mm/sec. This is a conservative limit on
which waves can escape detection because velocities
this small are unlikely to be detectable for the immediate
forseeable future \soho .

\figure\figgwaves{Here we plot, as a function of linear
frequency ($\nu_{n\ell} = \omega_{n\ell}/2\pi$),
the maximum relative amplitude, $\drho/\rho$,
which can be acheived by a $g$-wave for any $r$
less than half a solar radius. The dots are obtained by
requiring that the radial velocity of the wave
equal 1 mm/sec at the solar surface. Each dot represents
a wave having a different value of $n$ and $\ell$($\equiv$L).
Thin solid lines are drawn to guide
the eye which connect points having the same
values of $\ell$, but differing values for $n$.
The horizontal dashed line corresponds to a
density wave whose maximum amplitude is 1\% of
the unperturbed mass density. An analysis of
waves using linearized equations breaks down
for points much above this line.  The
shaded area indicates which modes may be
subject to instability \gmod, and so have much larger
amplitudes. The thick short straight line corresponds to a
wave energy of $10^{36}$ erg, which has been argued \gmod\
to be an energy which these waves could easily aquire, provided
they really are linearily instable \gmod. Modes below this line
would, if they had this energy, exhibit surface velocities
above $1$ mm/sec, increasing their chances of detection.
We are mainly interested in overstable modes above the
line, since these could have acceptably large energies
while still having surface velocities which are small
enough to escape detection.}

Figure \figgwaves\ gives a representation of the amplitude
which various $g$-waves would have if their surface radial
velocity were 1 mm/sec. Every dot in this plot gives the
frequency, $\nu_{n\ell}$, and the maximum relative
amplitude, $\drho/\rho_0$, which is obtained in the
interval $r/R_\odot \in [0, 0.5]$, for a wave labelled
by a pair of quantum numbers, $n$ and $\ell$.
Lines have been drawn to guide the eye,
connecting points which share a common value
for $\ell$ ($\equiv$ L in the figure). Clearly the figure shows
that $g$-waves can exist having any amplitude and
still escape detection, although any points in the
figure which appear much above the horizontal
dashed line representing a 1\% density fluctuation
should be considered unreliable due to the breakdown
of the linearized hydrodynamic equations we use.

Fig.~\figgwaves\ also shows that $g$-waves
having large $\ell$ and small $n$ are the ones which
can be largest deep in the solar interior without
much disturbing the solar surface. Large $\ell$ is
preferred because of two of the properties
outlined in the previous sections: ($i$) $g$-modes
having large $\ell$ tend to have their maximum
amplitudes near $r/R_\odot \sim 0.3$, making
them largest near resonance, and $(ii)$
$g$-waves have a factor which falls off with radius
like $r^{-\ell}$, which, for large $\ell$,
tends to suppress their influence at the solar surface.

Large-amplitude $g$-waves could therefore easily
escape current detection. How large might these
waves reasonably be expected to be? An answer
to this requires a theory of how these waves
are excited and damped. Waves which have
an appreciable overlap with the sun's convective
zone can be both excited and damped --- with
an efficiency believed to be roughly independent of
$n$ --- due to their interactions with this
zone's turbulent dynamics. $p$-waves, and
$g$-waves which penetrate sufficiently far
into the convective zone, can be excited in this
way.

A different mechanism applies for those $g$-waves
which don't extend far enough beyond the
radiative zone (see \gmod\ and references therein).
These modes are believed to be
stimulated in the energy-producing core, since
there any local compression of the solar medium
slightly increases the nuclear reaction
rates, leading to higher temperatures and pressures
and so to still more compression.  In the absence
of radiative damping of the resulting wave,
this production mechanism would lead to
a runaway instability. Radiative damping, however,
has an efficiency which increases with $n^2$, and
is therefore least effective for $g$-waves
of small radial order.

\ref\nonlinearity{P. Kumar and J. Goodman, {\bf Ap. J.} in press, (1996).}

Depending on whether excitation or damping
is most effective --- a subject of current
controversy --- $g$-waves having $n\le 3$
have been argued to have much larger amplitudes
than is expected for run-of-the-mill helioseismic waves.
In a linearized analysis these modes grow exponentially,
until they are so large that nonlinear effects
start to saturate their growth \nonlinearity .
An investigation of such nonperturbative
effects argues that these
low-$n$ modes could have energies as large as
$10^{35-37}$ erg \gmod\nonlinearity --- some 10 orders of magnitude
higher than the energies of other $g$- and $p$-waves.

The modes for which this instability may apply are
indicated in Figure \figgwaves\ by the shaded area.
In this shaded area, the thick, short straight line corresponds to a
wave energy of $10^{36}$ erg. Any mode which lies
below this line in the figure could not have an energy
as large as $10^{36}$ ergs without also having a
radial surface velocity larger than $1$ mm/sec, increasing
their chances of detection. Conversely, modes which lie above
the thick black line have larger values for $\ell$, and so
are damped more strongly with increasing $r$. They
could therefore have energies as large as $10^{36}$
erg and still have surface velocities which are too small
to be detected.

Thus, large-amplitude $g$-waves might
exist in the sun, although some {\it caveats}
must apply. First, the runaway excitation of $n\le 3$
$g$-modes is controversial~\gmod .
Even should such a runaway excitation mechanism exist,
it would have to preferentially excite modes having
large $\ell$, since low-$\ell$ modes are unlikely to
have escaped detection if their energies become too large.
In what follows we ignore such doubts, and
restrict ourselves to computing the influence
such waves would have on neutrino oscillations.
We do so in order to see whether $g$-waves can affect
the observed neutrino spectrum, even under the most
extreme circumstances.

\section{Implications for Solar Neutrinos}

Given the helioseismic wave profiles and normalizations
of the previous section, we are now positioned to
compute their implications for neutrino evolution.
We start with a brief description of some of the
issues which arise with our numerical simulations,
and then move on to the presentation of our results.

\subsection{The Numerical Algorithm}

Our numerical simulations were performed in the following steps.

\item{1)}
We first generate the wave profile for the desired helioseismic wave.

\item{2)}
We compute the survival probability for a single electron neutrino
which propagates through this wave. In the interests of speed
of calculation this was done simply by using Parke's formula
(as derived by Parke --- without fluctuations) for the
survival probability.
We verified for a sample of waves
that this accurately reproduced a direct numerical
solution of the MSW Schr\"odinger equation for the flavour
degrees of freedom in the presence of the helioseismic wave.
Agreement was good, although large-amplitude waves could
cause neutrinos to pass through multiple resonances, and
so it was necessary to use the Parke formula as applied to
multiple resonance crossings in order to get agreement with
numerical results.

\item{3)}
Step (2) was repeated for an ensemble of
200 different density profiles, obtained
by choosing the phase of the wave to be a
random variable. The survival probability for
each density profile was then averaged over the ensemble.

\item{4)}
For comparison, the generalized Parke formula was also used
to compute the ensemble-averaged survival probability.

\ref\profiles{J.N. Bahcall and M.H. Pinsonneault, \rmp{67}{95}{781}.}

\figure\figavgprob{We plot here the MSW survival probabilities
averaged over neutrino production points in the sun.
The dotted, dash-dotted and dashed lines show the
survival probabilities when averaged
over the production points of pp, ${}^7$Be
and ${}^8$B neutrinos, respectively. This is to be contrasted
with the solid line, which gives the result if all neutrinos
are assumed to have been produced at the exact
center of the sun. The average over the neutrino-production
site is an important background to any fluctuation calculation,
because both tend to smear out the MSW survival-probability
curve.}

We record here a number of detailed issues which
better define our numerical procedure:

\topic{Averaging Neutrino Production Sites}
Although frequently ignored in theoretical calculations,
neutrinos are not created exactly at the solar centre, and
are instead produced throughout the solar core,
out to several tenths of a solar radius. Furthermore,
different nuclear reaction cycles produce neutrinos
in different parts of the core, since their rates depend
strongly on temperature. In our code we
have used the production distributions of the
Bahcall-Pinsoneault SSM~\profiles\ for the Beryllium,
Boron and the proton-proton reactions.

The result of such an average for standard MSW oscillations
without fluctuations, is shown in fig.~\figavgprob.
To generate this figure neutrinos are randomly produced
(using Peter LePage's VEGAS Monte-Carlo routine)
throughout the sun with probability given by the Bahcall-Pinsoneault
distribution functions.
For practical reasons, the simulations are done using
around $200$ neutrino creation sites. This is sufficient
to obtain a precision of around $1$\% in
the Monte-Carlo integration over the neutrino production volume.
For simplicity we use spherical
coordinates centred at the solar centre, with the positive
$z$-axis pointing toward the earth.

\topic{Neutrino Path Geometry}
Since neutrinos are created at different points within
the sun, they see different density profiles
as they travel toward the earth depending on their
initial creation point. In our calculations we evolve each
neutrino along a path parallel to the $z$-axis,
starting at its production point. This is the path the
neutrino would take towards the earth, if the earth
were infinitely far from the sun. It is a very good
approximation to the real path taken to the earth
because the maximum "inclination" of this path
is of order $\theta \sim R_{sun}/R_{sun-earth} \sim 0.001$.

\topic{Multiple Resonance Crossing}
Since neutrinos are created throughout the solar
centre, they may cross the resonance region more
than once. This can happen if the neutrino starts further
from the solar centre than is the resonance radius,
but heads back through the solar centre on its
way to the earth.  Such neutrinos
can cross the resonance density twice, in which case the
jump probability in the Parke (or generalized Parke)
formula should be replaced by
$P_J \rightarrow 2 P_J (1 - P_J) $.

Similarly, a neutrino passing through a large-amplitude
helioseismic wave can pass through resonance more
than twice. This can happen because the presence
of the wave can make the solar density oscillate
back and forth across the resonance density several
times. A similar generalization of the jump probability
is required in this case.

\topic{More Than One Wave}
A limitation of our simulations is our assumption
that only one helioseismic wave is excited. In reality
most modes are excited simultaneously, increasing
the fluctuations seen by an ensemble of neutrinos.
In principle, the effect of exciting many modes is
computed by directly including all such
modes into the numerical ensemble average. Our
crude estimate of how our results would change
is to use an `equivalent amplitude' for a single wave.
For instance, if $N$ waves were to contribute a
roughly equal amount to the change in the neutrino
survival probability, and suppose a 1\%
density fluctuation is required for any one of these
waves to have an observable effect. Then we
estimate that if all $N$ waves are present
together, then the same observable effect
would be produced if each wave only had
an amplitude of $(1/\sqrt{N})\% $.
This kind of scaling is justified within the
master-equation approach for Gaussian
fluctuations when each mode is uncorrelated
with all of the others\foot\colen{Notice that a random
superposition of the relevant g-waves will not
significantly decrease the involved correlation
length since they all peak in the same area within
the sun and have similar wavelengths.}.
\endtopic

\subsection{Numerical Results}

We now turn to a discussion of the results of these
numerical simulations. Although we have run simulations
of neutrino propagation through many helioseismic
waves, we focus here on those $g$-waves
having radial degree $n\le 3$, since these are the ones
that could have the largest amplitudes. Furthermore we
consider only those modes whose radial surface velocities
are smaller than 1 mm/sec, which corresponds to choosing
$\ell \gsim 13$, (see Fig.~\figgwaves ). These waves are
superimposed on a background density profile, which we
take to be the Bahcall-Pinsonneault~SSM electron density
profile \profiles .

\figure\resultplots{We plot here the
survival probability for electron neutrinos
propagating through the $n=1$, $\ell = 13$
$g$-wave mode. Here the various dashed lines
represent the predictions of the generalized
Parke formula for varying amplitude
fluctuations. The long-dashed line corresponds
to: $(\drho/\rho_0)_{\rm max} = 0.3\%$;
the short-dashed line: 1\% and the dot-dashed
line: 3\%.
For comparison, the solid and dotted curves represent
a direct Monte Carlo averaging over 200 density
profiles, as described in the text. These are
largely indistinguishable from the pure MSW
prediction, even for fluctuations as large in
amplitude as $(\drho/\rho_0)_{\rm max}
= 1$ (dotted line). For simplicity of comparison
this plot is performed
assuming all neutrinos are produced right at the
solar centre. }

\figure\resultplottwo{
Same as Fig.~\resultplots\ the only difference being
that here we let the neutrinos propagate through
an $n=3$, $\ell = 50$ $g$-wave mode.
Again the various dashed lines represent the predictions
of the generalized
Parke formula. The long-dashed line corresponds
to: $(\drho/\rho_0)_{\rm max} = 0.3\%$;
the short-dashed line: 1\% and the dot-dashed
lines: 3\% and 10\% respectively.
The solid (1\% or less) and dotted (3\% and 6\%)
curves, on the other hand, represent
a direct Monte Carlo averaging, as described in the text.
These are
largely indistinguishable from the pure MSW
prediction, even for fluctuations as large in
amplitude as 3\% (lower dotted line).
For simplicity of comparison
this plot is performed
assuming all neutrinos are produced right at the
solar centre. }

Figures \resultplots\ and \resultplottwo\ give a typical result for the
size of the change in the electron-neutrino survival
probability as a function of the maximum amplitude
of the fluctuation. The dashed family of curves
in these figures gives the generalized Parke prediction
for an $n=1$, $\ell = 13$ (Fig.~\resultplots) and
an $n=3$, $\ell=50$ (Fig.~\resultplottwo) $g$-wave having
maximum amplitudes of
$0.3\%, 1\%, 3\%$ and $10\%$ within the radial
interval $r/R_\odot \in [0,0.5]$. These curves
differ strongly from the MSW prediction
throughout the resonance region, in agreement
with the earlier, preliminary estimates of Refs.~\noisy,
\noisytwo, \noisythree\ and \cliffdenis.
By contrast,
the solid and dotted curves give the result of a direct
numerical ensemble average, for the various amplitudes
indicated, as described in the previous sections.
These latter curves are indistinguishable
from the MSW prediction for all values of $E/\Delta m^2$,
even for fluctuations as large as 3\% in
size.

At first sight the small size of the predicted effect
is surprising, given the fact that the $g$-wave of
interest is strongest for $r/R_\odot \in [0.1-0.3]$,
which is exactly where ${}^7$Be neutrinos pass through
resonance. This small result is the consequence of
the long-wavelength of the wave in question,
which is of order $R_\odot/10 \sim 7 \times 10^4$ km.
This long wavelength is much longer than the
typical neutrino oscillation lengths, taking it well
outside the domain of validity of the master-equation
approach. This puts us in the large correlation length limit of
Fig.~\figcomp, where deviation from the MSW
effect becomes negligible.

Notice that the main difference between Figure~\resultplottwo\
($n=3$, $\ell=50$ g-wave) as opposed to
Figure~\resultplots\ ($n=1$, $\ell=13$ g-wave) is a decrease in
size of the effect when computed using the generalized Parke formula
and an increase in size when employing the
direct numerical ensemble average, the latter leading to sizeable effects
when the wave amplitude exceeds $3\%$ (dotted curves in
Fig.~\resultplottwo ). This difference is due to a decrease in
correlation length when going from a mode of radial order $n=1$
to $n=3$. Two remarks are in order when comparing these figures.
First, the difference in $\ell$ does is {\it not} responsible for the
difference between these figures, since the neutrinos in these simulations
were taken to originate in the solar centre. As a result, a larger value
for $\ell$ doesn't much change the correlation length as seen by
the outgoing neutrino. Second, the comparably large effect shown
for the dotted ($\drho/\rho_0 = 6\%$) curve in Figure~\resultplottwo\
is an artifact of starting all  neutrinos at the solar centre,
since those produced further out miss the first upward `bump' in the
radial density profile, leaving the remaining downward and upward
bumps to largely cancel. Once averaged over neutrino production
sites, the effect on neutrino oscillations is drastically reduced,
so much that evan a $g$-mode having $\drho/\rho_0 = 1$
generates no detectable deviation from the MSW spectrum.

With this result it is clear that helioseismic waves
can have no observable effect on neutrino oscillations.
Even if a $g$-wave were imagined having a maximum amplitude
as large as 1\%,\foot\noway{Notice that even though such
modes don't produce detectable surface motion, and although
their time-dependence prevents them being ruled out by
seismic inversion methods, waves having this amplitude
have enormously high energies of about $\ss 10^{46}$ erg ! }
the suppression of its influence on neutrino
oscillations due to its long wavelength would even so
cause it to negligibly alter neutrino oscillations. This
disagrees with all previous predictions because these
are based on the master-equation approach, which fails
for waves having wavelengths as large as these.

\section{Conclusions}

We close with a summary of our conclusions, and a few
speculations on how our results may affect other, more
exotic, types of solar-neutrino solutions.

\subsection{Summary of Results}

In this paper we have carefully modelled neutrino passage
through helioseismic waves, and have shown that helioseismic
waves induce no observable consequence on MSW oscillations.
Conversely, if MSW oscillations should be observed in the next generation
of neutrino telescopes, then experimental agreement with
MSW predictions provides no constraint on the size
of density fluctuations in the resonance region that can
be probed using linearized methods.

There are four main reasons underlying our conclusion.
These are:

\item{1)} Neutrino oscillations appear to be sensitive to short-range
fluctuations that are a few percent in amplitude in the immediate
vicinity of any MSW resonance. With the possible exception of
low-$n$, large-$\ell$ $g$-modes, however, helioseismic waves
are many orders of magnitude too small at resonance
depths within the sun to affect neutrino oscillations.

\item{2)} Low-$n$, large-$\ell$ $g$-waves can have much larger
 amplitudes because they are potentially subject to an instability
to which other modes are insensitive. Unfortunately, such modes
have wavelengths along the neutrino line of sight which
are very large compared to neutrino oscillation lengths. The
same is true of random combinations of many such modes,
since all low-$n$, large-$\ell$ $g$-modes tend to peak at the
same depths within the sun, and to have similar wavelengths.
This puts these waves beyond the domain of applicability of all
previous calculations, and turns out to preclude
even these potentially runaway modes from affecting
neutrino oscillations.

\item{3)}
By providing the first calculation which goes beyond
the small-correlation-length approximation, we have found that
long-wavelength waves produce effects which are
additionally suppressed compared to previous estimates.
We therefore find an additional requirement if
fluctuations are to significantly influence neutrino
oscillations: the correlation length of the fluctuation
should be roughly the same size as the perturbation length scale,
which we define to be the length scale at whcih the master-equation
approach fails. This length scale is typically of the same
order of magnitude as the neutrino oscillation length.

\item{4)} Finally, any effect whose signature is a smearing of
the adiabatic dip in the MSW survival probability profile, can potentially
be masked by the similar smearing which happens purely within
the ordinary MSW framework when the average over the
neutrino-production site is taken (see fig.~\figavgprob ).

We conclude that density fluctuations in the sun
are unlikely to modify the small angle MSW neutrino spectrum.
Most of the known fluctuations in the sun
are too small at the neutrino resonance point to
affect the predicted neutrino spectrum. The same is not
true for helioseismic waves, but these nevertheless also
neglibly influence neutrino oscillations.
We know of no other plausible mechanism which
can influence MSW oscillations.

\subsection{Implications for nonMSW Solutions to the
Solar Neutrino Problem}

One of the major conlusions to emerge from the study of
the influence of density perturbations on neutrino propagation
is that the survival probability is sensitive almost exclusively to
fluctuations at or close to the resonance \cliffdenis.
MSW resonances are not influenced by most
solar fluctuations simply because most fluctuations
don't take place near the MSW resonance.

The present study adds to this the additional information
that the correlation length should be comparable to the
oscillation length, if a fluctuation is to appreciably influence
neutrino oscillations.

\ref\mms{A. Cisneros, {\it Astro. \& Space Sci.} {\bf 10} (1971) 87;\bk
M.B. Voloshin, M.I. Vysotskii and L.B. Okun, {\it Sov. Phys. JETP}
{\bf 64} (1986) 446;\bk
C.S. Lim and W.J. Marciano, \prd{37}{88}{1368};\bk
E.K. Akhmedov, \plb{213}{88}{64}. }

Although MSW oscillations always occur well away from
the convective zone, and so well away from its associated
density fluctuations, this is not true for some other proposals
to solve the solar neutrino problem. In particular,
it is not true for
magnetic-moment solutions, in which neutrinos
are conjectured to posses a small magnetic moment
which induces flavour oscillations in the presence of large
magnetic fields \mms. In this scenario neutrinos
can also undergo resonant conversions in the presence
of matter, with the important difference that
some of these resonances can occur within
the convective zone. The survival probability could therefore
{\it in principle} be much more strongly modified by
the strong fluctuations in the magnetic field and electron density
within the convective zone. The conclusions of this paper
show that in order to produce an effect any such fluctuations
must also have correlation lengths which
are comparable to the typical oscillation lengths for
these neutrinos. The potential
interplay between convective-zone
fluctuations and resonant magnetic-moment
oscillations would bear closer study.

\bigskip
\centerline{\bf Acknowledgements}
\bigskip

We would like to acknowledge helpful correspondence
with Professors Joergen Christensen-Dalsgaard,
Pawan Kumar and John Bahcall concerning helioseismology.
Our research is
financially supported by NSERC of Canada and FCAR
du Qu\'ebec.

\listrefs

\figurecaptions



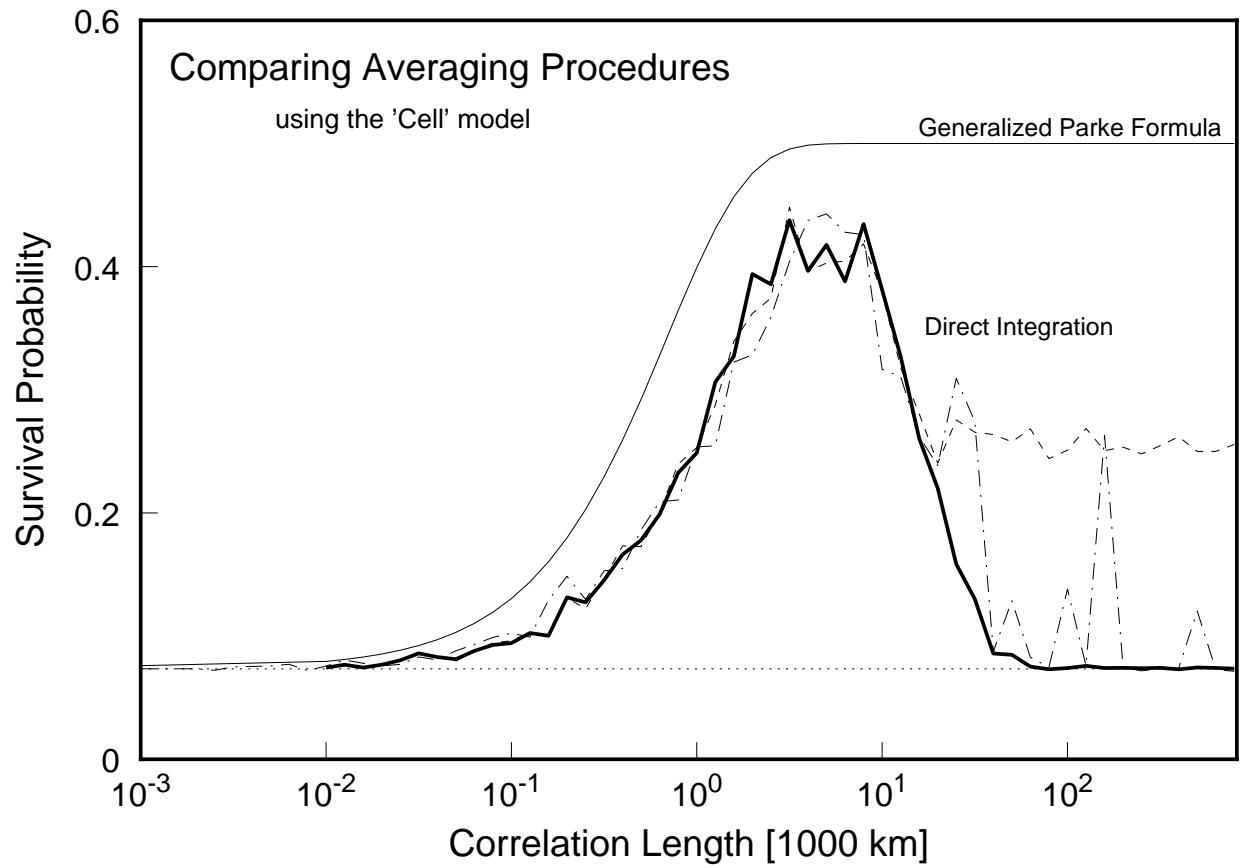
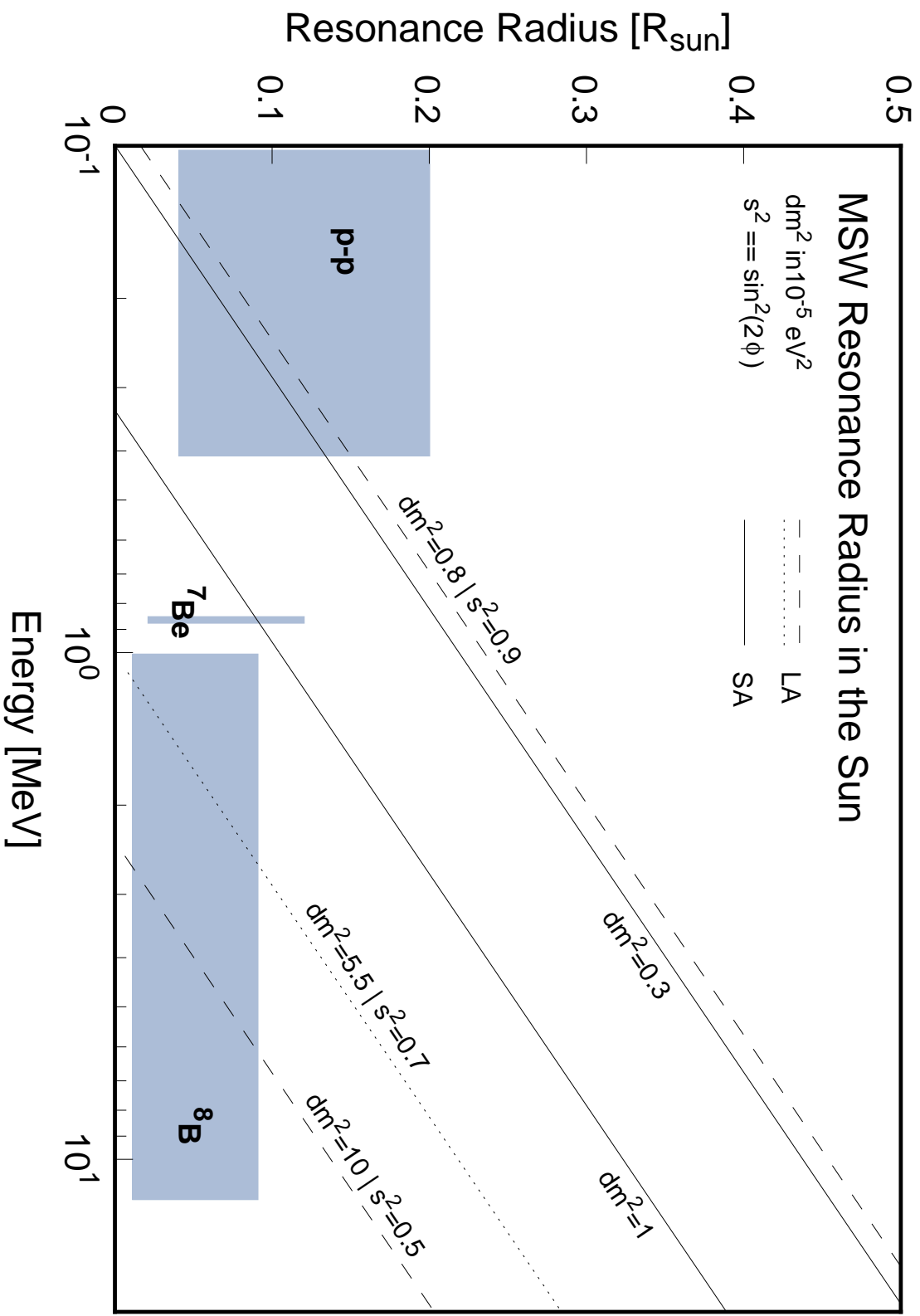
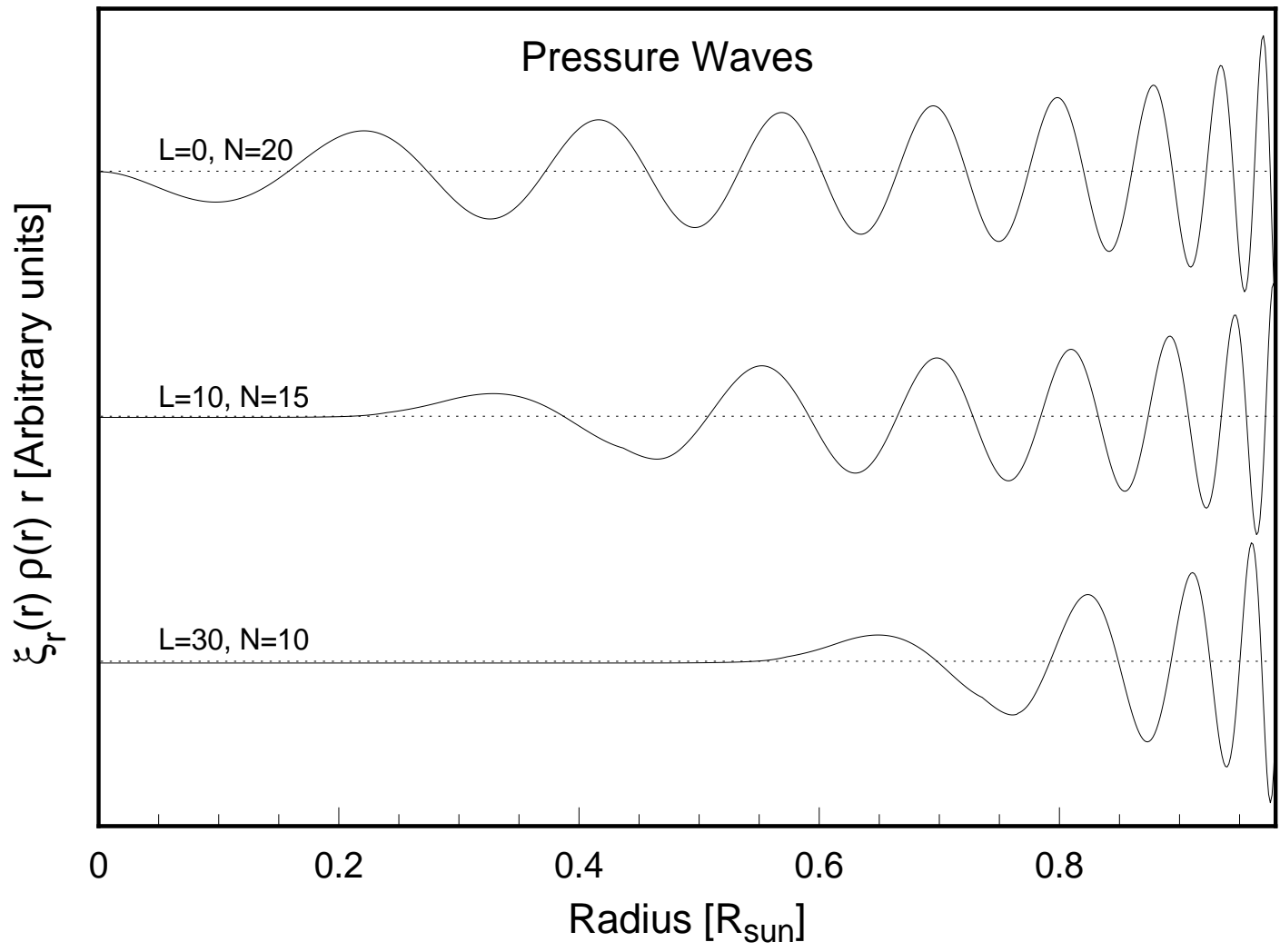
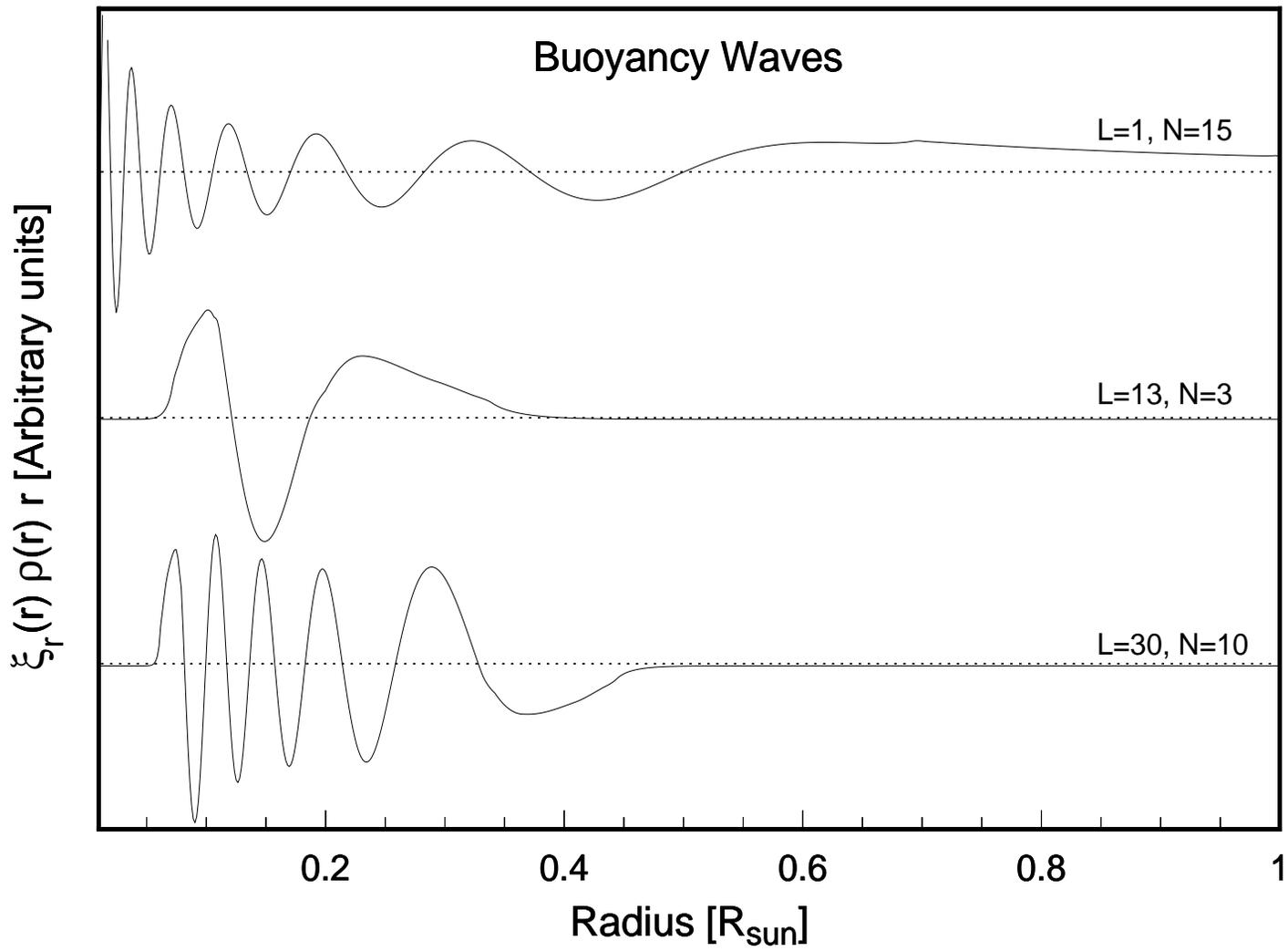
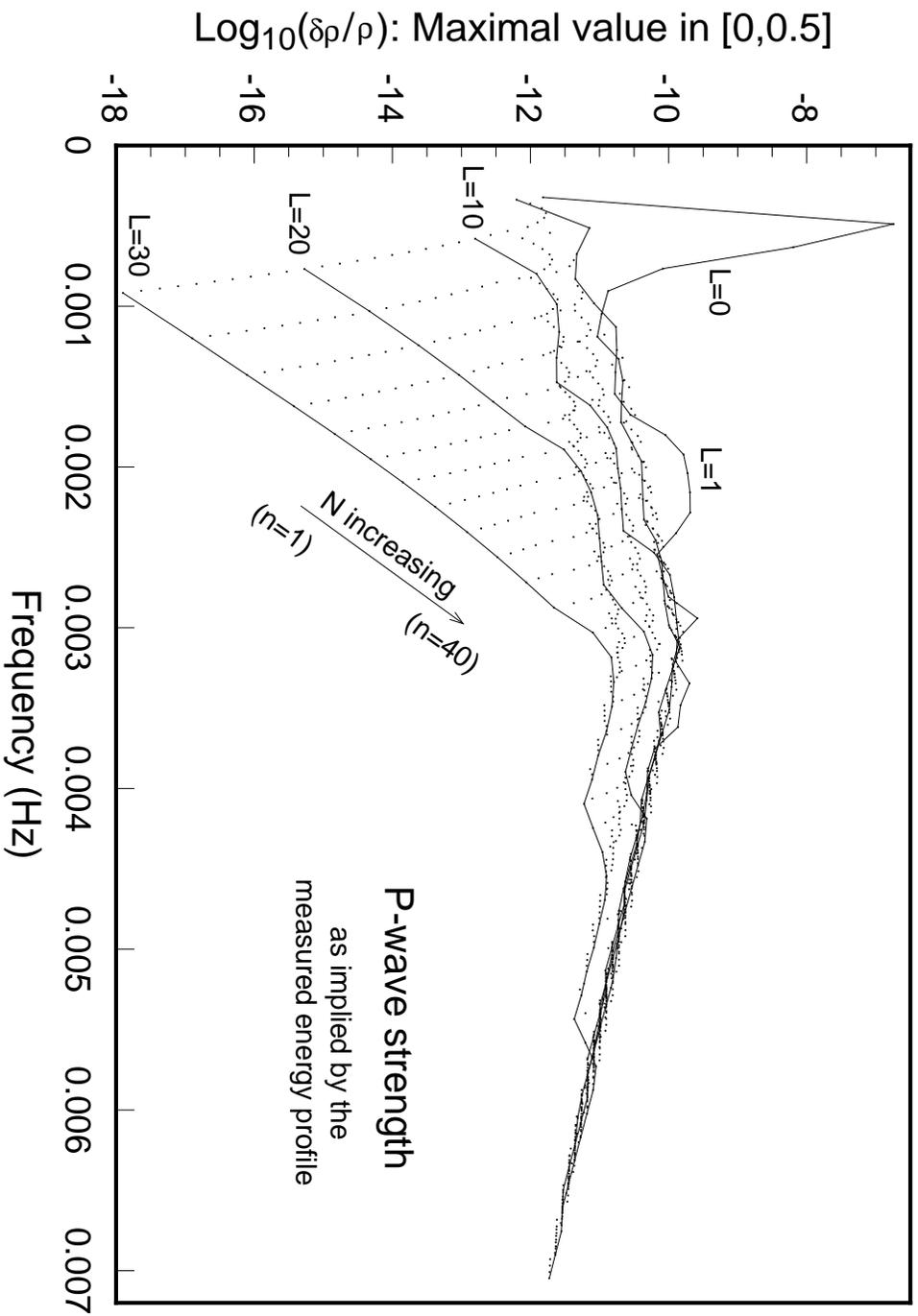
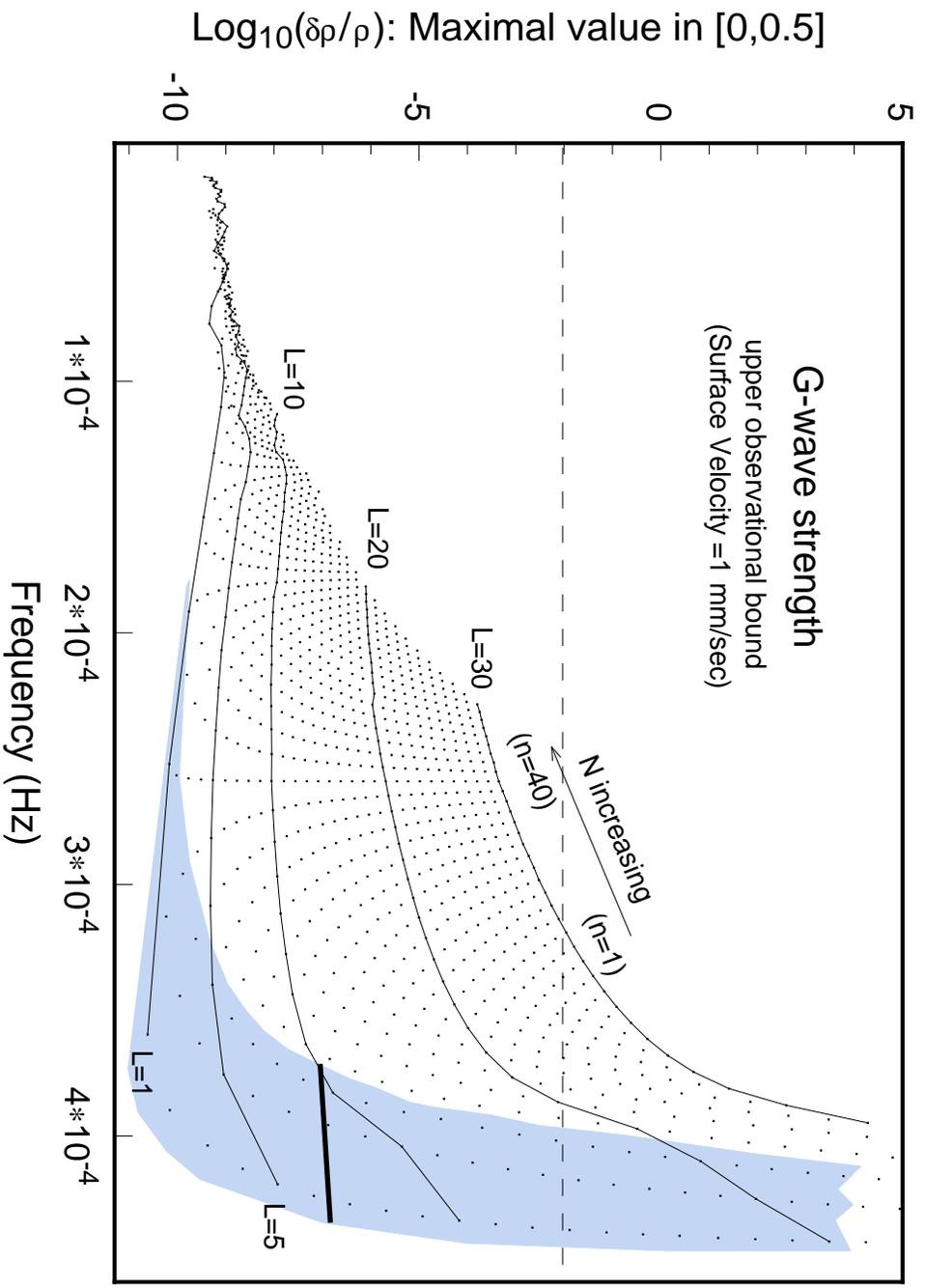
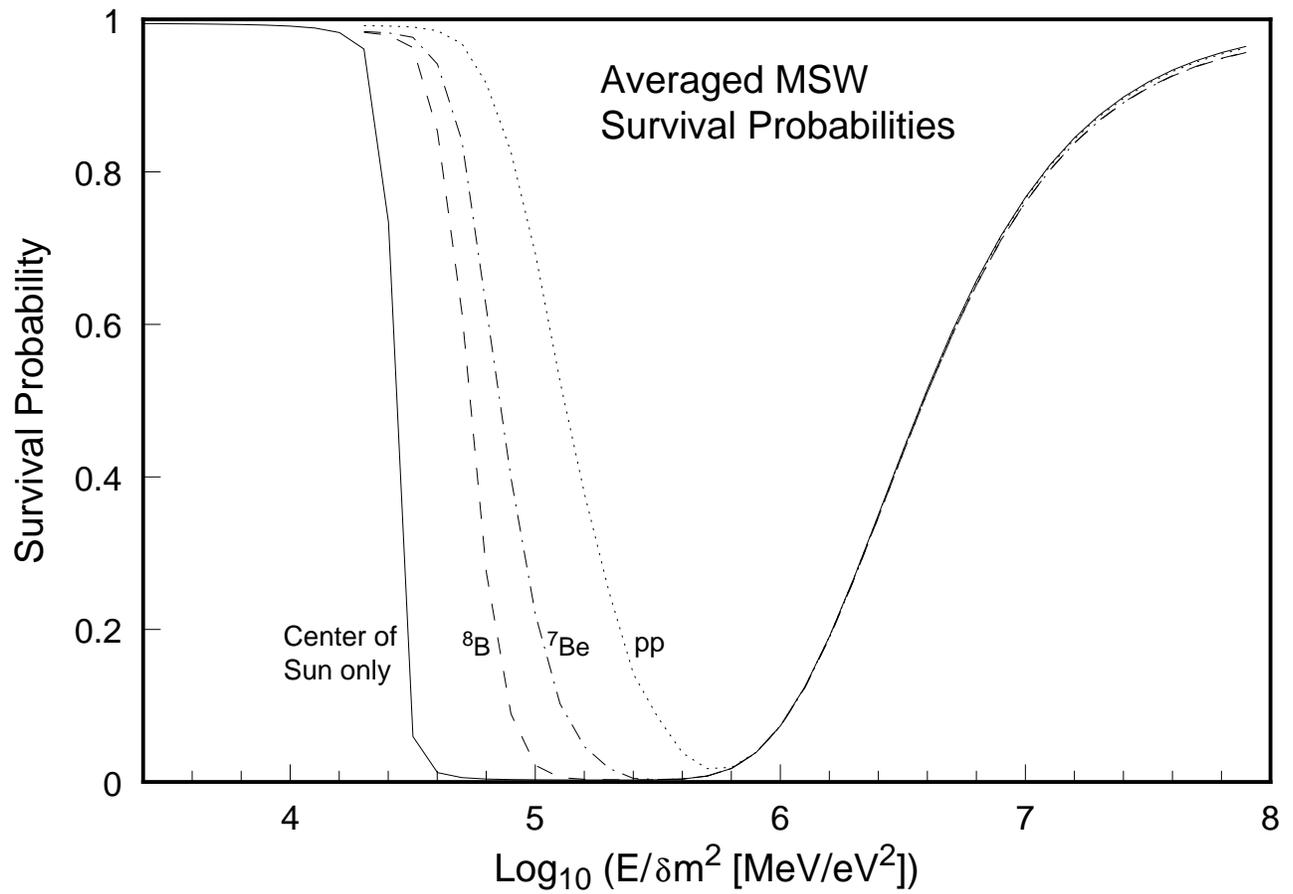
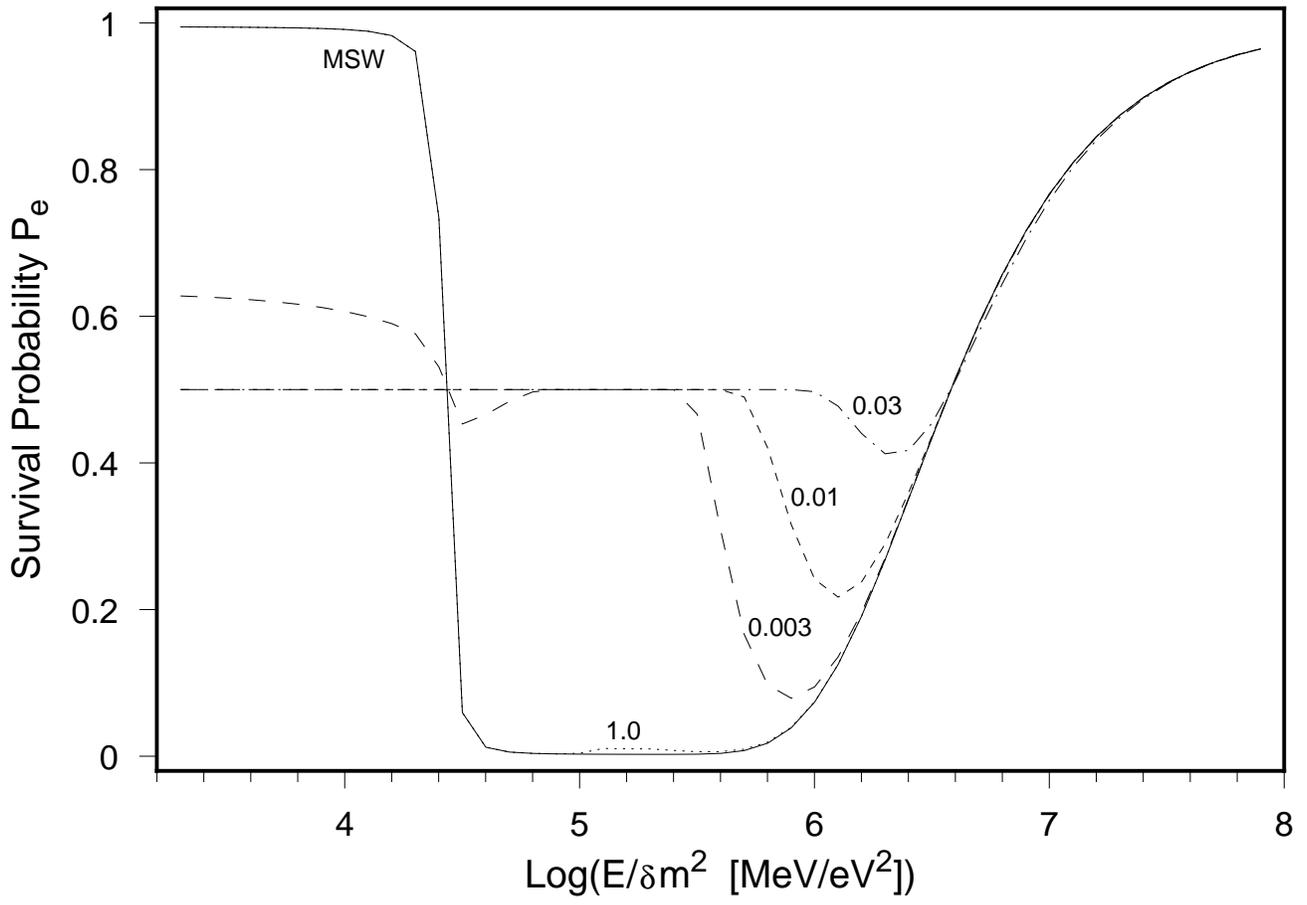
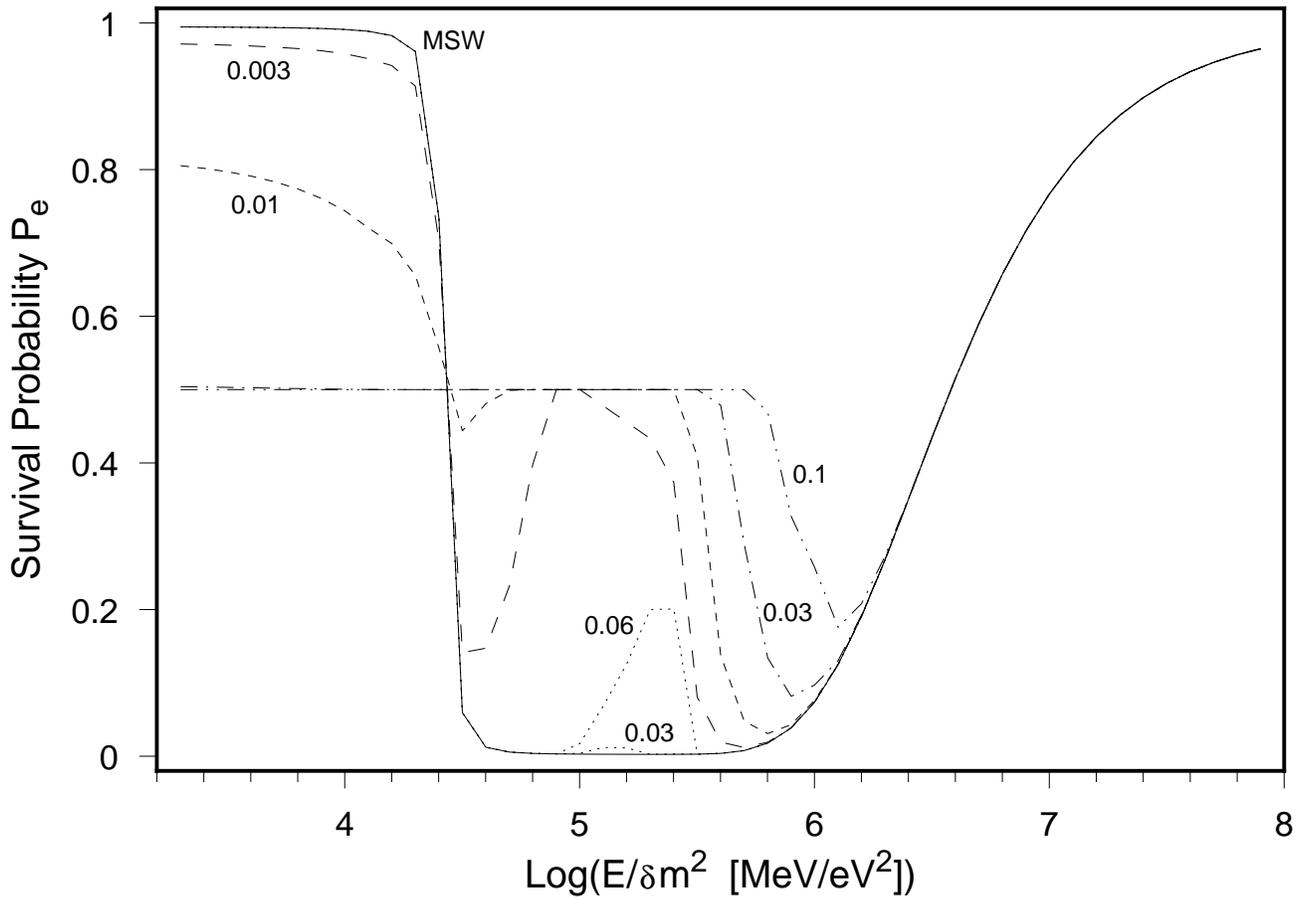
\bye